\documentclass[conference]{IEEEtran}

\usepackage{cite}
\usepackage{amsmath,amssymb,amsfonts}
\usepackage{graphicx}
\usepackage{textcomp}
\usepackage{xcolor}
\usepackage{booktabs}
\usepackage{multirow}
\usepackage{url}
\usepackage{xcolor}
\usepackage{xcolor}
\usepackage{xspace}

\newcommand{\drop}[1]{\textcolor{red}{#1}}
\newcommand{\gain}[1]{\textcolor{green!60!black}{#1}}

\newcommand{\tool}{TRAIL\xspace}
\newcommand{\phead}[1]{\vspace{0.4em}\noindent\textbf{#1}}
\usepackage[most]{tcolorbox}
\begin{document}

\title{TRAIL: Towards Effective Log Parsing for Rare Logs through Template-Preserving Augmentation}

\author{
\IEEEauthorblockN{Sepideh Hodaeian}
\IEEEauthorblockA{
University of Alberta\\
Edmonton, Alberta, Canada\\
\texttt{sephod@ualberta.ca}
}
\and
\IEEEauthorblockN{Zhenhao Li}
\IEEEauthorblockA{
York University\\
Toronto, Ontario, Canada\\
\texttt{lzh9410@gmail.com}
}
\and
\IEEEauthorblockN{An Ran Chen}
\IEEEauthorblockA{
University of Alberta\\
Edmonton, Alberta, Canada\\
\texttt{anran6@ualberta.ca}
}
}

\maketitle

\begin{abstract}
Log parsing is essential for system log analysis because it supports tasks
such as debugging, monitoring, and anomaly detection by transforming
unstructured log messages into structured log templates. However, real-world log datasets exhibit highly imbalanced, long-tailed distributions, where a small number of frequent templates dominate while many rare templates appear only a few times. 
This imbalance causes evaluation results to be overly optimistic by allowing frequent templates to dominate benchmark metrics, while poor performance on rare yet operationally important events remains largely hidden.

In this paper, we investigate the prevalence and impact of rare log groups, defined as log groups with fewer than five instances. Our empirical study on the widely used Loghub-2.0 benchmark shows that rare log groups account for nearly 20\% of all templates but less than 0.01\% of log messages. 
Because they contain only a handful of instances, all evaluated parsers experience substantial performance degradation on these groups.

To address this challenge, we propose \tool, a log parsing framework that improves template inference for rare log groups through template-preserving augmentation. 
\tool enriches rare log groups with template-consistent log messages before template inference. The additional structural evidence helps distinguish static tokens from dynamic variables.
Our experimental results show that \tool improves parsing accuracy on rare log groups by 19\% over the strongest baseline while maintaining competitive performance on complete datasets.
We further show that the proposed augmentation strategy generalizes across different LLM backbones and consistently improves existing LLM-based parsers without modifying their core architectures.

\end{abstract}


\section{Introduction}

\begin{figure}[!t]
    \centering
    \includegraphics[width=\columnwidth]{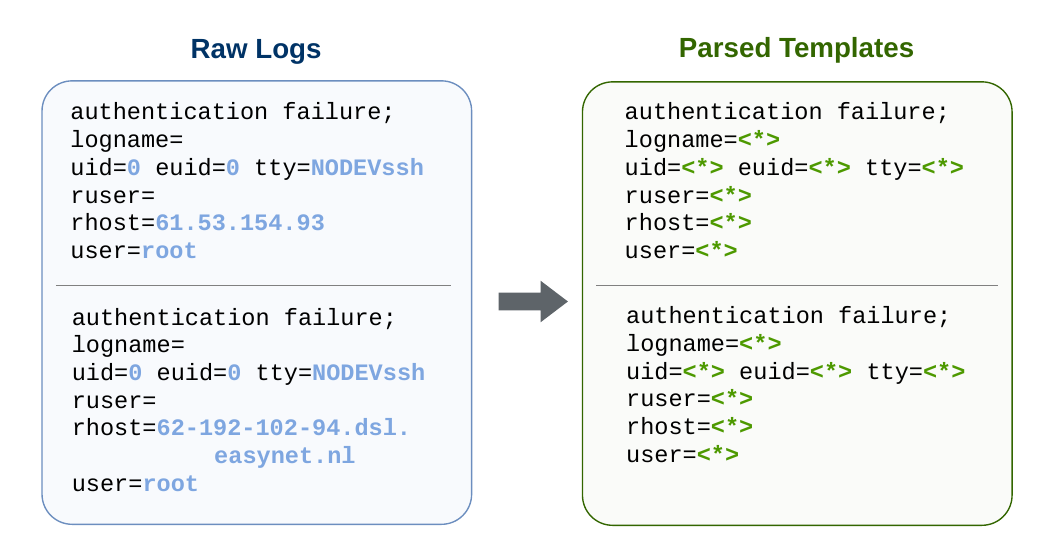}
    \caption{Example of log parsing from the Linux dataset.}
    \label{fig:logexample}
\end{figure}

Software systems continuously generate large volumes of logs during execution~\cite{zhu2019tools,Logram2020}. These logs record runtime events, system states, warnings, and failures, providing valuable information about software behavior. Log analysis~\cite{He2021AutomatedLogAnalysisSurvey,Jiang2025L4,He2022MicrosoftLogAnalysis,He2016SystemLogAnalysis} plays a fundamental role in many software engineering tasks, including software debugging~\cite{xu2009detecting,Yuan2010Sherlog,chen2019empirical,chen2021pathidea}, system monitoring~\cite{Wang2021QuickPeek,Chen2019LoadTests,candido2021logbased,razavi2008pattern}, anomaly detection~\cite{Chen2019LoadTests,he2017towards,He2016SystemLogAnalysis,sun2026improving,Wang2023LogOnline}, failure prediction~\cite{das2020failure,lim2008log,zheng2009system,zhang2016automated}, and root cause analysis~\cite{lu2017logbased,zawawy2010log,suriadi2012root,mariani2008automated}.
However, before logs can support these downstream tasks, they must first be transformed into a structured representation~\cite{zhu2015learning,shin2021theoretical,Li2023VariableAware}. Because raw log messages contain both static event descriptions and dynamic runtime values, they are inherently semi-structured. Log parsing 
transforms these messages into structured log templates 
by identifying static tokens shared across log messages while extracting dynamic runtime values into parameters. As illustrated in Figure~\ref{fig:logexample}, log parsing converts two raw log messages with different runtime values into a common log template.
This enables downstream log analysis to operate on structured log events rather than individual log messages.

To improve the effectiveness of log analysis, a wide range of log parsing techniques have been proposed over the past decade.
Traditional parsers, such as Drain~\cite{Drain2017} and Spell~\cite{Spell2019}, rely on heuristic rules and token-based matching to efficiently infer templates. More recently, large language model (LLM)-based parsers, including LibreLog~\cite{LibreLog2024}, LILAC~\cite{LILAC2024}, and EFParser~\cite{wang2026small}, have further improved parsing performance by leveraging semantic reasoning to identify template structures across diverse logging formats. 

The effectiveness of these parsers depends on the availability of representative log messages for template inference. During normal execution, software systems often generate the same log messages, whereas templates corresponding to failures, exceptional events, or uncommon execution paths typically occur only a handful of times. Consequently, these templates provide much less structural evidence for template inference.
Table~\ref{tab:example} compares a frequent and a rare template from the Mac dataset to illustrate this challenge.
The frequent template \texttt{E113} accounts for 7.7\% of all log messages, and LibreLog correctly parses every instance (parsing accuracy = 1.00). In contrast, the rare template \texttt{E164} appears only three times, accounting for less than 0.01\% of all log messages, reducing the parsing accuracy to 0.00. This example highlights how limited structural evidence can substantially degrade parser performance on rare templates.
Since rare templates often correspond to failures, exceptional events, or uncommon execution paths, incorrectly parsing them can hide operationally important system behaviors from downstream analysis.

\begin{table}[!t]
\centering
\caption{LibreLog's parsing performance on frequent and rare group samples in the Mac dataset. PA denotes the parsing accuracy.}
\label{tab:example}
\resizebox{\columnwidth}{!}{
\begin{tabular}{p{6.5cm}|ccr}
\toprule
\multicolumn{1}{c|}{\textbf{Expected Template}} & \textbf{GroupId} & \textbf{Ratio} & \textbf{PA} \\
\midrule

\texttt{<*>}: scheduler\_evaluate\_activity told me to run this job; however, the start time is not for \texttt{<*>} seconds. Ignoring.
& \multirow{2}{*}{E113} & \multirow{2}{*}{7.7\%} & \multirow{2}{*}{1.00} \\

\midrule

handle will sleep auth and shield windows: Ordering out authw \texttt{<*>} (\texttt{<*>}), shield \texttt{<*>} (\texttt{<*>}) (lock state: \texttt{<*>}).
& \multirow{2}{*}{E164} & \multirow{2}{*}{$<$0.01\%} & \multirow{2}{*}{0.00} \\

\bottomrule
\end{tabular}
}
\end{table}

Although recent work has shown that parser performance varies with template frequency~\cite{jiang2024large}, it neither formally defines rare log groups nor investigates their prevalence in real-world datasets. To address this gap, we conduct an empirical study on the Loghub-2.0 benchmark~\cite{he2020loghub}. Our study reveals that rare log groups account for a substantial proportion of log templates despite representing only a tiny fraction of log messages. We further observe that existing parsers struggle on these groups, which suggests that limited structural evidence remains a fundamental challenge for accurate template inference.

Motivated by these findings, we propose \tool, a log parsing framework that reconstructs the missing evidence required for template inference in rare log groups. \tool achieves this by generating additional log messages that preserve the original log structure while introducing different variable values. These augmented examples help parsers better distinguish static tokens from dynamic variables. Because this process is performed before parsing, it can be integrated with existing parsers without modifying their underlying algorithms.

In summary, this paper makes the following contributions:

\begin{itemize}
    \item We conduct the first study on rare log groups in Loghub-2.0, showing that they are common, challenging to parse, and often overlooked by existing evaluation metrics.
    \item We propose \tool, a log parsing technique that improves template inference for rare log groups through template-preserving augmentation.
    \item Our extensive evaluation on over 50 million log messages from Loghub-2.0 demonstrates that \tool improves parsing accuracy on rare log groups by at least 19\% over state-of-the-art log parsers while maintaining strong performance and robustness on complete datasets.
    \item We show that template-preserving augmentation is a general technique that improves existing LLM-based parsers by up to 41\% in parsing accuracy and remains effective across multiple LLM backbones.
\end{itemize}

\noindent Our replication package is publicly available.\footnote{Available at: \url{https://anonymous.4open.science/r/TRAIL-parser-6A5F/}}

\begin{figure*}[t]
\centering

\begin{minipage}{0.329\textwidth}
\centering
\includegraphics[width=\linewidth]{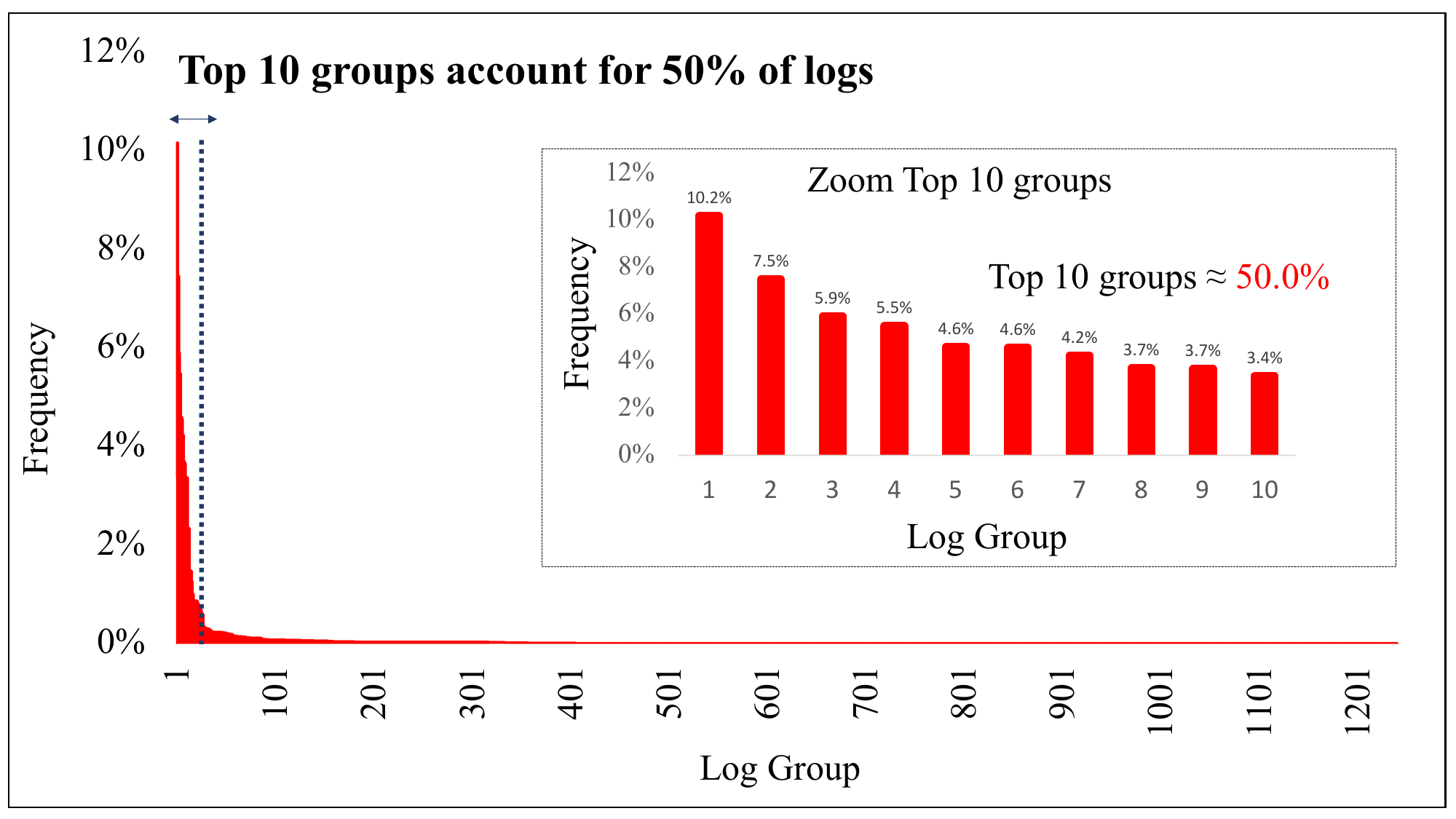}
\\(a) Thunderbird
\end{minipage}
\hfill
\begin{minipage}{0.329\textwidth}
\centering
\includegraphics[width=\linewidth]{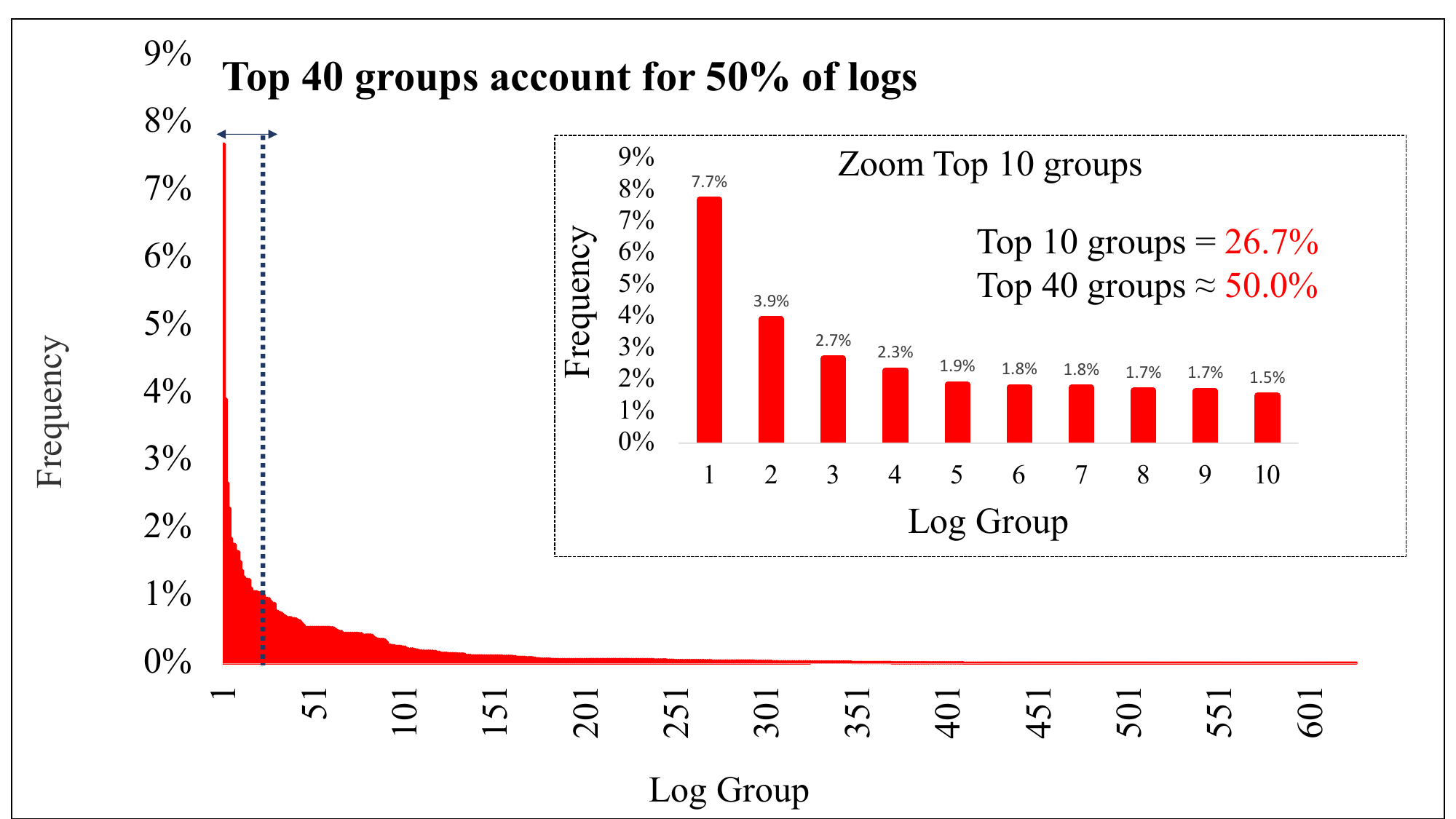}
\\(b) Mac
\end{minipage}
\hfill
\begin{minipage}{0.329\textwidth}
\centering
\includegraphics[width=\linewidth]{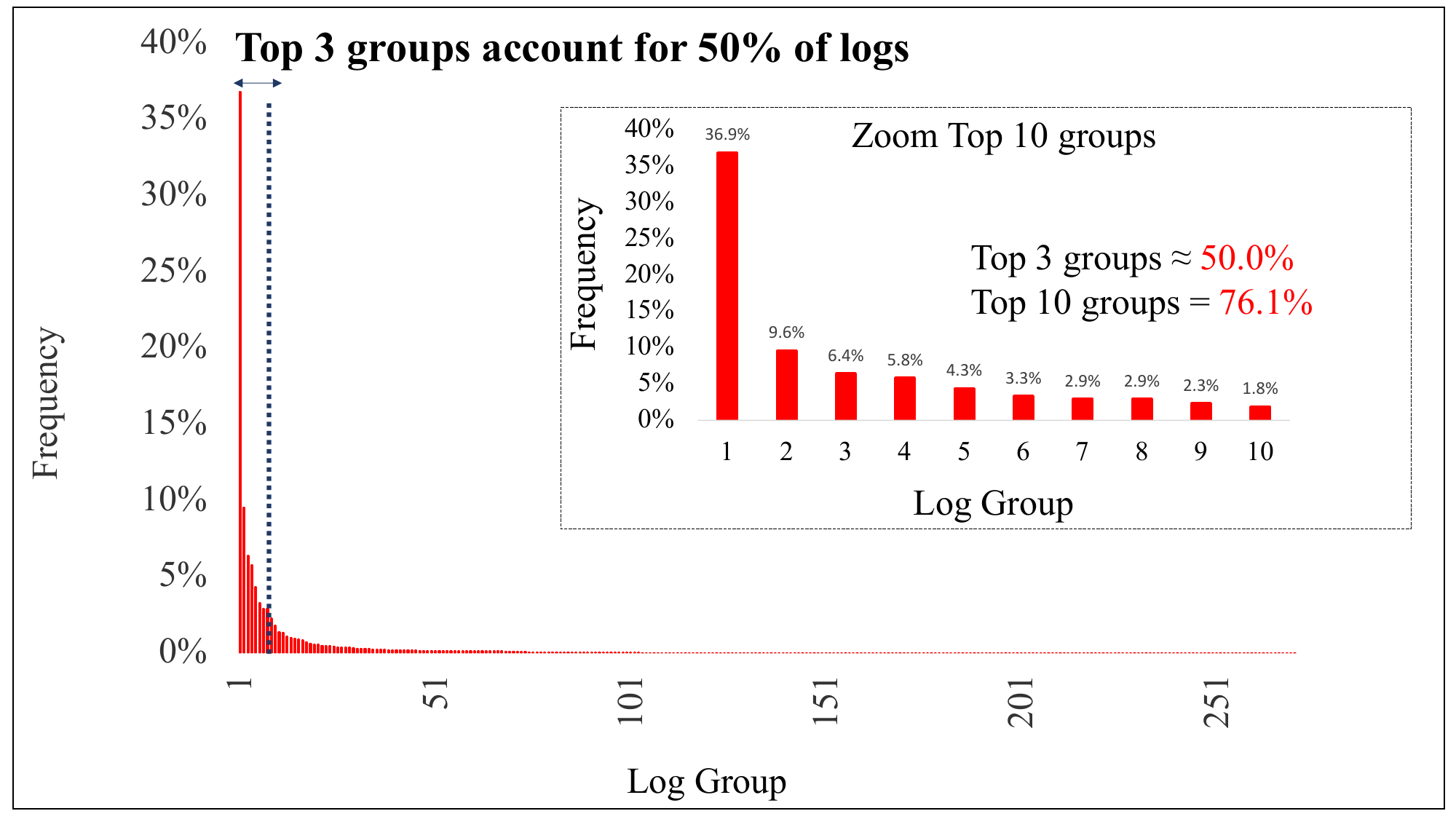}
\\(c) BGL
\end{minipage}

\vspace{0.3cm}

\begin{minipage}{0.329\textwidth}
\centering
\includegraphics[width=\linewidth]{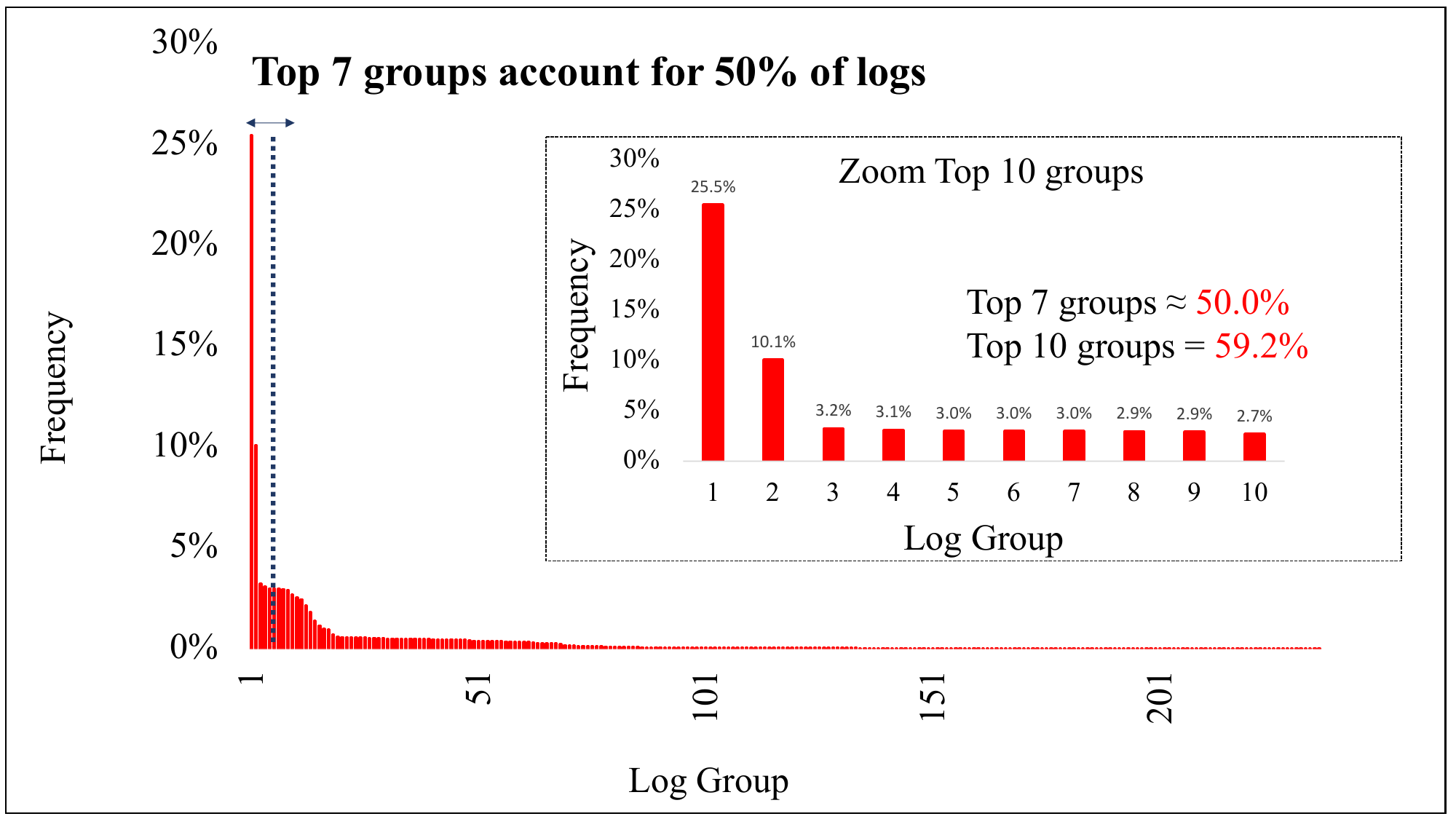}
\\(d) Hadoop
\end{minipage}
\hfill
\begin{minipage}{0.329\textwidth}
\centering
\includegraphics[width=\linewidth]{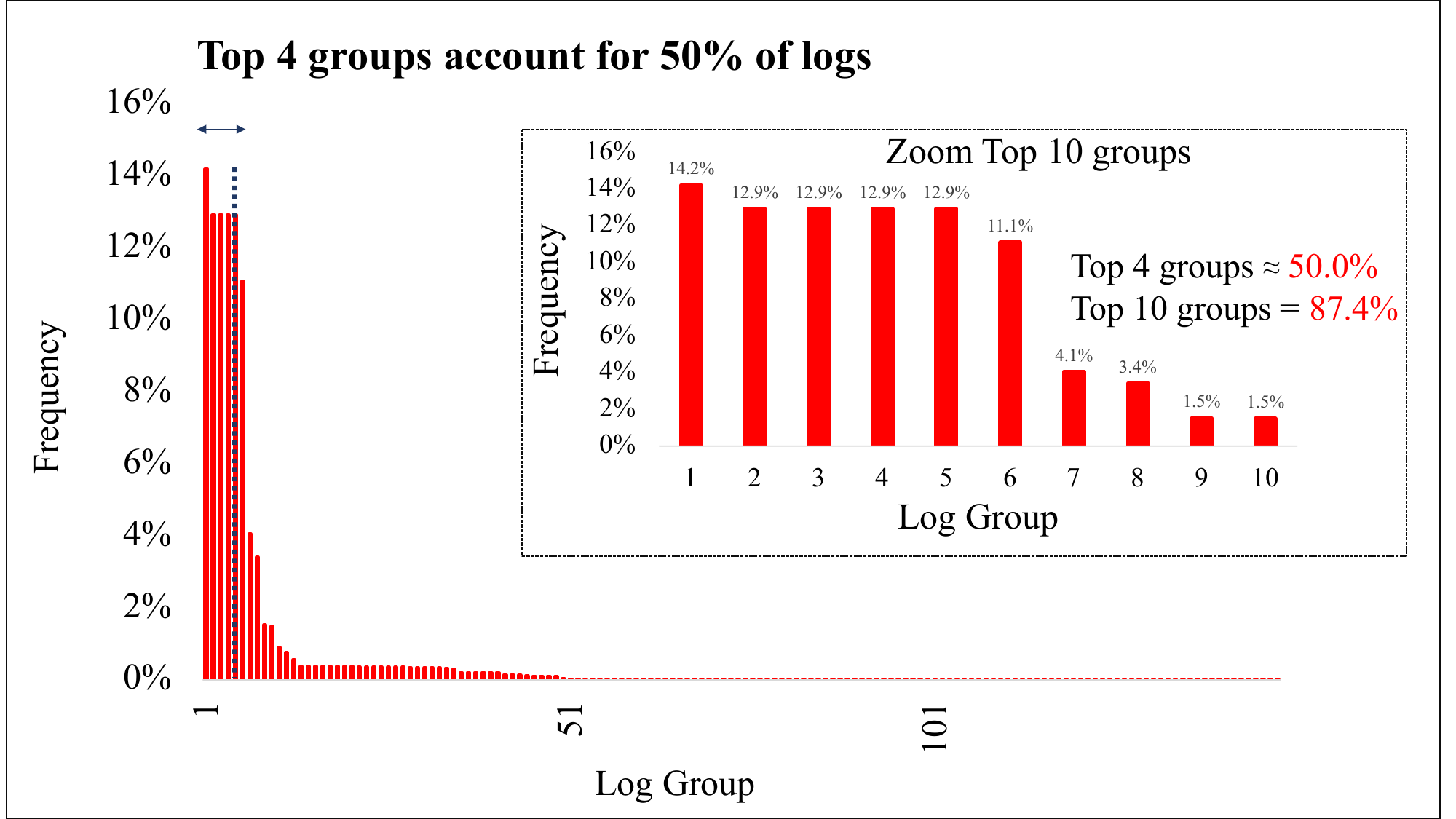}
\\(e) HealthApp
\end{minipage}
\hfill
\begin{minipage}{0.329\textwidth}
\centering
\includegraphics[width=\linewidth]{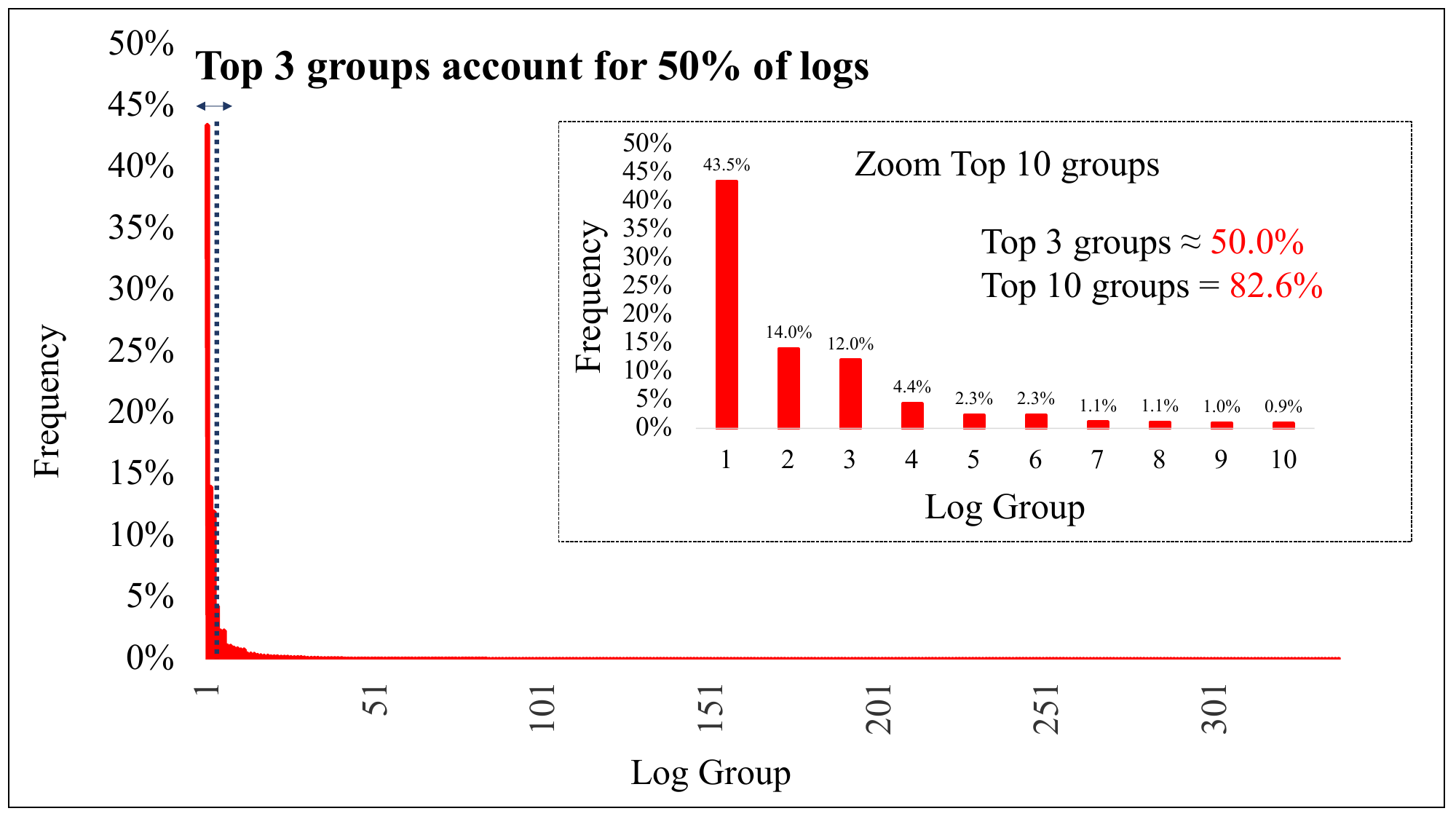}
\\(f) Linux
\end{minipage}

\caption{Distribution of log messages across log groups in Loghub-2.0.}
\label{fig:distribution}

\end{figure*}
\section{Empirical Study on Rare Logs}\label{section2}

In this section, we first investigate the prevalence of rare log groups in real-world systems and examine their impact on log parser performance. Following our definition, a rare log group contains fewer than five log instances. We analyze all 14 datasets in Loghub-2.0 to answer two questions: 1) \textit{How common are rare log groups?} 2) \textit{How well do existing parsers handle them?}

\subsection{Identifying and Characterizing Rare Logs}

\begin{table}[!t]
  \centering
  \caption{Statistics of Loghub-2.0 datasets and rare log groups}
  \label{tab:Table0_fixed}
  \resizebox{\columnwidth}{!}{

  \begin{tabular}{l|rrrr}
    \toprule
    \textbf{Dataset} &
    \textbf{\# Logs} &
    \textbf{\# Rare Logs} &
    \textbf{\# Groups} &
    \textbf{\# Rare Groups} \\
    \midrule
    Mac         & 100,315         & 331 (0.03\%) & 626     & 175 (27\%) \\
    Thunderbird & 16,601,746      & 682 ($<$0.01\%) & 1,242   & 316 (25\%) \\
    Linux       & 23,922          & 154 ($<$0.01\%) & 332 & 65 (19\%) \\
    HPC         & 429,989         & 26  ($<$0.01\%) & 74 & 13 (17\%) \\
    OpenSSH     & 638,948         & 18  ($<$0.01\%) & 38 & 6 (15\%) \\
    Hadoop      & 179,994         & 78  ($<$0.01\%) & 236 & 35 (14\%) \\
    Zookeeper   & 74,274          & 18  ($<$0.01\%) & 89 & 7 (13\%) \\
    Spark       & 16,075,118      & 41  ($<$0.01\%) & 236 & 23 (9\%) \\
    BGL         & 4,631,262       & 53  ($<$0.01\%) & 320 & 26 (8\%) \\
    HDFS        & 11,167,741      & 10   ($<$0.01\%) & 46 & 4 (8\%) \\
    HealthApp   & 212,395         & 27  ($<$0.01\%) & 156 & 12 (7\%) \\
    Apache      & 51,979          & 2   ($<$0.01\%) & 29 & 2 (7\%) \\
    OpenStack   & 207,633         & 6   ($<$0.01\%) & 49 & 3 (6\%) \\
    Proxifier   & 21,321          & 0 (0\%) & 11& 0 (0\%) \\
    \midrule
     Total       &50,419,938       & 1,446 ($<$0.01\%)   &3,484   & 687 (20\%)\\
    \bottomrule
  \end{tabular}
  }
  
\end{table}

\phead{Definition of rare logs.}
A log group consists of log messages that share the same ground-truth template. 
We define a \textbf{rare log group} as a log group containing fewer than five log messages, and refer to log messages belonging to rare log groups as \textbf{rare logs}.
We use an absolute threshold because template frequencies vary greatly across datasets, which makes percentage-based thresholds dependent on dataset size.
We empirically select five as the threshold because it represents the smallest group size that typically provides sufficient structural evidence for reliable template inference. Groups with fewer than five instances often lack enough variation for parsers to distinguish static tokens from dynamic variables, while groups with five or more instances generally become substantially easier to parse. Therefore, we focus on the most challenging cases where limited structural evidence is likely to affect parser performance.

\phead{Studied datasets.}
We conduct our experiments on the Loghub-2.0  benchmark provided by He et al.~\cite{he2020loghub}, a widely used benchmark for evaluating log parsing techniques. Loghub-2.0 contains log datasets collected from 14 open-source systems spanning diverse domains, including distributed systems, supercomputer platforms, and server-side applications. Due to differences in system scale, workload, and logging practices, the included datasets exhibit substantially different log distributions. Compared with the earlier Loghub-1.0~\cite{he2020loghub} benchmark, Loghub-2.0 significantly expands both the scale and diversity of the evaluation data. In total, the dataset comprises more than 50 million log messages.

\phead{Prevalence of rare log groups.}
We begin by examining how frequently rare logs occur in practice.
Table~\ref{tab:Table0_fixed} presents the results.
We observe that rare log groups are present in 13 out of the 14 studied datasets. Across all datasets, rare log groups constitute nearly 20\% of all log groups, with the proportion ranging from 6\% to 27\%, except for the Proxifier dataset, which contains no rare log group. Despite this relatively high percentage of rare log groups, the number of log messages belonging to these groups is extremely small, accounting for less than 0.01\% of the total logs. This indicates a substantial imbalance in log distribution.
A particularly striking example is the Thunderbird dataset, where only 682 out of 16,601,746 log messages are classified as rare, yet these messages correspond to about 25\% of all log groups. These results suggest that many templates have limited occurrences but together represent a significant portion of the template space.

\begin{tcolorbox}{\textbf{Observation 1: }}{Although rare log groups account for nearly 20\% of all templates, they contribute less than 0.01\% of log messages.}
\end{tcolorbox}
\phead{Long-tailed distribution analysis.}
To better understand this imbalance, we analyze the distribution of log messages across log groups. Figure~\ref{fig:distribution} shows the six representative datasets in Loghub-2.0, while the complete set of distributions is available in our replication package~\cite{TRAIL_repo}.
All datasets exhibit a pronounced long-tail distribution.
For example, in the Thunderbird dataset, as illustrated in Figure~\ref{fig:distribution} (a), the top 10 log groups (out of 1,242) alone account for more than 50\% of all log messages, while the remaining groups occur infrequently. Although rare log groups contribute only a small fraction of log messages, they may capture important system behaviors. For instance, the template ``\texttt{<*> more authentication failure; logname=<*> uid=<*> euid=<*> tty=<*> ruser=<*> rhost=<*>}'' appears only twice in the Thunderbird dataset, yet indicates repeated authentication failures that may signal a severe security anomaly.
These findings show that Loghub-2.0 exhibits a pronounced long-tail distribution, where a small number of log groups account for most log messages while many groups occur only a few times. Such imbalance leaves rare log groups with insufficient structural evidence for reliable template inference. Moreover, because evaluation metrics are typically dominated by the large number of messages from frequent groups, they may overestimate parser performance while masking failures on rare logs. To examine whether these effects occur in practice, we next evaluate the performance of state-of-the-art log parsers on 
rare log groups and frequent log groups (i.e., all log groups containing five or more log instances) separately.

\begin{table*}[!t]
    \centering
    \caption{Performance comparison of log parsing approaches on frequent and rare log groups.}
    
    \label{tab:parser_compariso}
    \resizebox{0.9\textwidth}{!}{
    
    \begin{tabular}{l|l|llll|llll}
        \toprule
        \multirow{2}{*}{\textbf{Parser}} & \multirow{2}{*}{\textbf{Category}} & \multicolumn{4}{c|}{\textbf{Frequent Log Groups}} & \multicolumn{4}{c}{\textbf{Rare Log Groups}} \\
         && PA & GA & FTA & FGA & PA (\gain{↑}/\drop{↓}) & GA (\gain{↑}/\drop{↓}) & FTA (\gain{↑}/\drop{↓}) & FGA (\gain{↑}/\drop{↓}) \\
        \hline
        AEL      & Rule-based &  0.38 & 0.79 & 0.27 & 0.58 & 0.32 (15\%\drop{↓}) & 0.95 (20\%\gain{↑})& 0.30 (11\%\gain{↑}) & 0.89 (53\%\gain{↑}) \\
        Drain    &Tree-based & 0.44 & 0.86 & 0.32 & 0.58 & 0.21 (52\%\drop{↓}) & 0.98 (13\%\gain{↑}) & 0.27 (15\%\drop{↓}) & 0.98 (68\%\gain{↑}) \\
        LILAC-8shots    & LLM-based (Llama-3-8)
            &0.69 & 0.86 & 0.56 & 0.72 & 0.39 (43\%\drop{↓}) & 0.78 (9\%\drop{↓}) & 0.43 (23\%\drop{↓}) & 0.83 (15\%\gain{↑}) \\
        LibreLog & LLM-based (Llama-3-8) &0.82 & 0.89 & 0.47 & 0.66 & 0.60 (26\%\drop{↓})  & 0.93 (4\%\gain{↑}) & 0.57 (21\%\gain{↑})  & 0.83 (25\%\gain{↑}) \\
        EFParser & LLM-based (Llama-3-8) & 0.80 & 0.87 & 0.73 & 0.90 & 0.67 (16\%\drop{↓}) & 0.98 (12\%\gain{↑}) & 0.67 (8\%\drop{↓}) & 0.98 (8\%\gain{↑})\\
        \bottomrule
        \addlinespace
        \multicolumn{10}{p{0.9\textwidth}}{\footnotesize
        \textbf{Note:} Percentages in the rare log groups columns denote the relative performance change with respect to the corresponding parser on frequent log groups. \gain{↑} and \drop{↓} denote performance improvement and degradation, respectively.
        }
    \end{tabular}
    }
\end{table*}

\begin{tcolorbox}{\textbf{Observation 2: }}{The widely-used Loghub-2.0 benchmark exhibits a strong long-tail distribution, where a few log groups generate most messages while many rare groups occur only a few times but may correspond to critical system events.}
\end{tcolorbox}

\subsection{Performance of State-of-the-art Log Parsers on Rare Logs}\label{sota_performance}

\phead{Baselines.}
We evaluate five representative log parsers covering major families of existing approaches, including rule-based (AEL~\cite{AEL2008}), tree-based (Drain~\cite{Drain2017}), and LLM-based parsers (LILAC~\cite{LILAC2024}, LibreLog~\cite{LibreLog2024}, and EFParser~\cite{wang2026small}).
These parsers were selected because they represent diverse template inference mechanisms and have been widely adopted as baselines in recent log parsing studies~\cite{he2020loghub,zhu2019tools,jiang2024large}. In particular, the three LLM-based parsers leverage different state-of-the-art template inference strategies, which allows us to evaluate modern parsing techniques under imbalanced log distributions.

\begin{itemize}
    
    \item \textbf{Drain.} A tree-based log parser that incrementally organizes log messages into a fixed-depth tree based on token similarity.

    \item \textbf{AEL.} A rule-based method that groups logs using heuristic clustering and token frequency analysis.

    \item \textbf{LILAC.} An LLM-based parser that performs template inference through in-context learning with demonstration examples.

    \item \textbf{LibreLog.} An unsupervised LLM-based method that retrieves logs with diverse variable values to guide template inference.

    \item \textbf{EFParser.} An unsupervised LLM-based parser that leverages adaptive cache updates and template correction for both accurate and efficient log parsing.
    
\end{itemize}

\phead{Experimental environment.}
We conduct all experiments on a Linux-based machine equipped with an Intel Core i9-7920X CPU (12 cores) at 2.90 GHz, 125 GB of RAM, and four NVIDIA GeForce RTX 2080 Ti GPUs (11 GB memory each).
To ensure fair comparison, we evaluate all baseline methods under the same execution environment using their default configurations.
For LLM-based parsing, we use Llama-3-8B with temperature set to 0 for deterministic inference. We choose Llama-3-8B because it is a widely used open-source model with a good balance between performance and computational cost.
We reproduce all baseline implementations from their public repositories using the original or recommended parameters. 
Following prior work~\cite{huang2025no,wang2026small}, we replace LILAC's in-context learning module with an 8-shot prompting configuration (denoted as LILAC-8shots) to enable fair comparison.

\phead{Evaluation metrics.}
We evaluate log parsing performance using four metrics covering both the log and template levels:
\begin{itemize}
    \item \textbf{Parsing Accuracy (PA).}
    Ratio of correctly parsed log messages over the total number of log messages. This metric~\cite{Logram2020} evaluates whether the parser assigns each log message to the same template as the ground truth.

    \item \textbf{Grouping Accuracy (GA).}
    Ratio of correctly grouped log messages over the total number of log messages. This metric~\cite{zhu2019tools} evaluates whether log messages are assigned to the correct log groups.

    \item \textbf{F1-score of Template Accuracy (FTA).}
    Harmonic mean of precision and recall of template accuracy. This metric~\cite{jiang2024large} evaluates whether the inferred log templates exactly match the ground-truth templates. A predicted template is correct if and only if all log messages assigned to it originate from a single ground-truth template and the static tokens of the predicted template exactly match those of the ground truth. This metric penalizes over-generalized or incorrectly extracted templates that may otherwise achieve high log-level accuracy.
    
    \item \textbf{F1-score of Group Accuracy (FGA).}
    Harmonic mean of precision and recall of group accuracy. This metric~\cite{jiang2024large} measures how accurately a log parser groups log messages into correct templates. It is computed based on the precision and recall of grouping accuracy. 

\end{itemize}


\phead{Results.}
Table \ref{tab:parser_compariso} compares the performance of different log parsers on frequent and rare log groups. 

\noindent\textbf{Rare log groups remain challenging for all existing log parsers, regardless of the underlying parsing strategy.}
Across all parser families, including rule-based, tree-based, and LLM-based approaches, parsing performance degrades substantially on rare log groups. 
For instance, the tree-based parser Drain experiences the largest reduction in PA at 52\%, while the LLM-based parser LILAC-8shots suffers a 43\% decrease. Even the strongest LLM-based parsers, LibreLog and EFParser, are not immune, with performance drops of 26\% and 16\%, respectively.
This finding suggests that the challenge is not tied to a particular parsing strategy, but rather to the limited structural evidence in rare groups. When only a handful of examples are available, parsers have fewer opportunities to distinguish static tokens from dynamic variables.
Although LLM-based parsers generally outperform traditional approaches on frequent log groups, their advantage becomes much smaller on rare log groups.
For example, LILAC-8shots performs comparably to, or worse than, the traditional parsers on multiple metrics. These findings suggest that stronger semantic reasoning alone is insufficient when parsers lack enough structural evidence for reliable template inference.

\noindent\textbf{Grouping rare log messages appears to be considerably easier than accurately recovering their templates.}
For instance, Drain achieves a PA of only 0.21 on rare logs while GA remains strong. Similar trends are observed for AEL, LibreLog and EFParser. These results indicate that existing parsers can generally identify which rare log messages belong together.
One possible explanation is that rare log groups contain only a small number of log messages, which makes grouping relatively straightforward, while accurately identifying static and dynamic tokens still requires sufficient structural variation.
This discrepancy also highlights a limitation of grouping-based evaluation metrics. Because GA and FGA reward correct clustering without considering whether the extracted templates are accurate, they may overestimate parser robustness on rare log groups.

Taken together, these findings reveal that the primary challenge of rare log parsing is not grouping similar messages, but recovering accurate templates under limited structural evidence. Existing parsers can often recognize that rare logs belong together, yet they struggle to infer which tokens are static and which represent dynamic parameters. This observation motivates approaches that explicitly enrich structural evidence before template inference, which forms the central idea behind \tool.

\begin{tcolorbox}
\textbf{Observation 3:} Rare log groups remain challenging for both traditional and LLM-based parsers. Our findings suggest that limited structural evidence is the primary cause of parsing errors on rare logs.
\end{tcolorbox}

\section{Methodology}

\subsection{Overview}

In this section, we present \tool, a log parsing framework designed to improve template inference under limited structural evidence. The key idea behind \tool is to reconstruct missing structural evidence before template inference. Rather than inferring templates directly from a limited number of log messages, TRAIL enriches underrepresented log groups with diverse, template-consistent examples. These examples help distinguish static tokens from dynamic variables.

\begin{figure}[!t]
    \centering
    \includegraphics[width=\columnwidth, keepaspectratio]{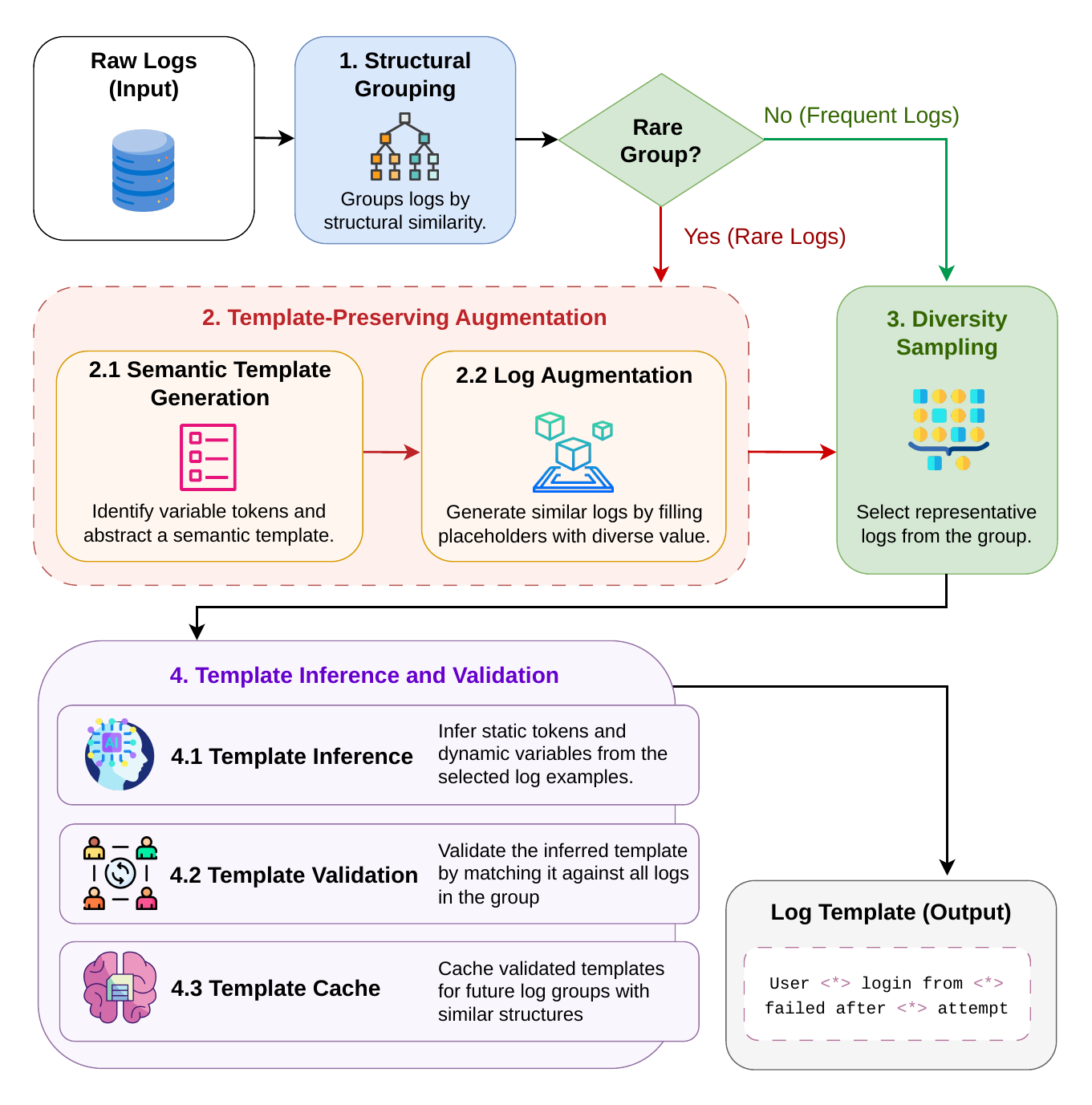}
    \caption{Overview of TRAIL.}
    \label{fig:TRAIL}
\end{figure}

Figure~\ref{fig:TRAIL} presents the overall workflow. The framework consists of four main stages: \textbf{(1) Structural Grouping}, which partitions log messages into structurally similar groups. \textbf{(2) Template-Preserving Augmentation}, which reconstructs missing structural evidence through semantic template generation and log augmentation. \textbf{(3) Diversity Sampling}, which selects representative examples that maximize structural variation. \textbf{(4) Template Inference and Validation}, which infers, validates, and caches log templates.

\subsection{Structural Grouping}
We first partition log messages into structurally similar groups. The objective of this stage is to identify groups that may suffer from limited structural evidence. 
To construct these groups, we adopt a lightweight grouping strategy inspired by Drain~\cite{Drain2017}. 
Each log message is tokenized and normalized by replacing numeric values with a generic placeholder. The normalized logs are then organized using a fixed-depth parsing tree based on message length and leading tokens. Logs with similar token structures are assigned to the same branch, while new branches are created when no suitable match exists. After all log messages have been processed, each leaf node corresponds to a group of structurally similar log messages.

After a log group is formed, we classify it based on its size. Following the definition in Section~\ref{section2}, groups with fewer than five log instances are considered rare log groups. 
Only rare groups undergo template-preserving augmentation to enrich their structural evidence, whereas frequent groups proceed directly to diversity sampling. The representative log instances selected by diversity sampling are then used for template inference and validation.

\subsection{Template-Preserving Augmentation}
Rare log groups provide insufficient evidence for template inference. With few examples, it is difficult to determine which tokens are static and which represent variables. To address this challenge, we enrich these groups through template-preserving augmentation before diversity sampling.
The augmentation process consists of two steps: \textit{Semantic Template Generation} and \textit{Log Augmentation}.

\begin{figure}[!t]
    \centering
    \includegraphics[width=0.99\columnwidth]{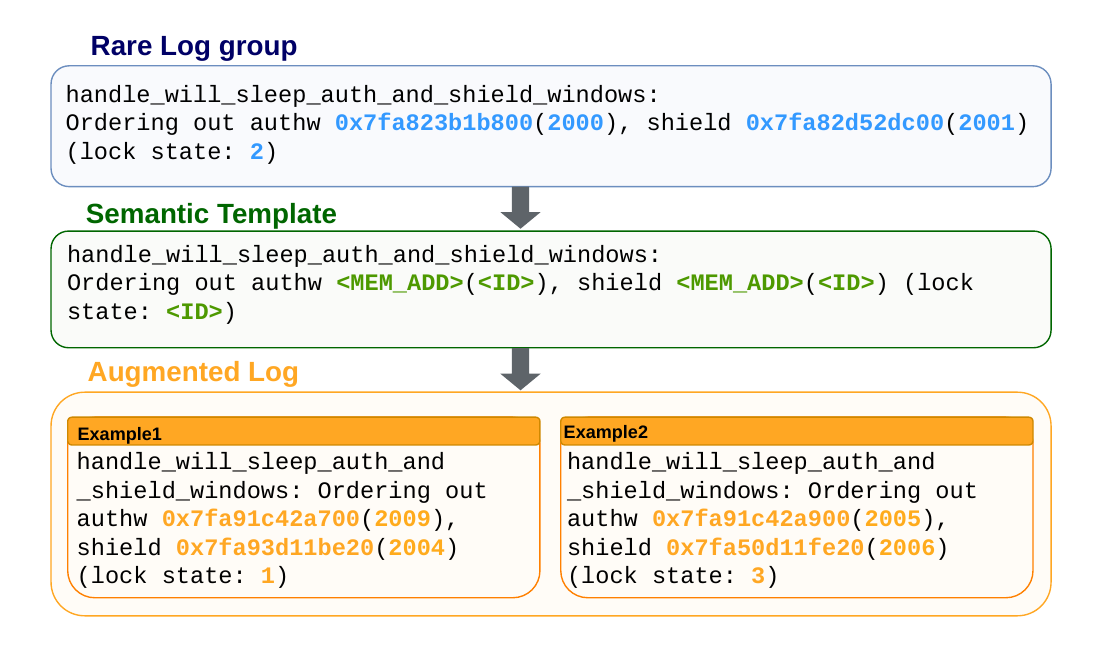}
    \caption{Example of Template-Preserving Augmentation for a rare log group from the Mac dataset.}
    \label{fig:augexample}
\end{figure}

Figure~\ref{fig:augexample} illustrates the process using a rare log group containing a single log message. We first derive a semantic template that identifies common semantic categories within a log message while preserving the underlying log structure. We then generate alternative values for each semantic category and substitute them back into the semantic template to produce multiple template-consistent log instances.
Although the variable values differ, all generated logs share the same underlying template, which provides additional structural evidence for template inference.
For illustration purposes, Figure 4 shows only two augmented examples. In practice, we generate five augmented instances for each rare log group.


\noindent\underline{\textit{Semantic Template Generation.}}
We first derive a semantic template that identifies the role of variable tokens within a log group.
Before semantic template generation, we apply a small set of tokenization constraints to preserve structured entities, such as IP addresses, file paths, URLs, and host:port values, as single tokens.
We then apply 15 lightweight regular-expression recognizers covering the most common structured entities found in logs, including memory addresses, process identifiers, timestamps, IP addresses, file paths, URLs, and user names. Matched values are replaced with their corresponding semantic placeholders that represent their categories (e.g., \texttt{<MEM\_ADD>} and \texttt{<ID>}) to construct the semantic template. The recognizers are designed to capture only common structured entities rather than every possible variable type. Tokens that do not match any recognizer remain unchanged and are handled directly during template inference. The complete list of recognizers is available in our replication package.

\noindent\underline{\textit{Log Augmentation.}}
To provide additional structural evidence for template inference, we prompt an LLM to generate alternative values for each semantic category while preserving the surrounding log structure.
Rather than prompting an LLM to generate complete log messages, we prompt it to generate alternative values for each semantic placeholder. For each placeholder, we provide the original value as context and ask the model to generate alternatives from the same semantic category (e.g., replacing an IP address with another IP address or a memory address with another memory address). This ensures that only variable values change while the underlying template remains the same.
We then construct augmented logs by replacing the semantic placeholders with the generated values. As shown in Figure~\ref{fig:augexample}, the augmented logs preserve the original template while introducing additional variation in the dynamic variables. 
We generate five augmented log instances for each rare log group and combine them with the original log instances to form the candidate pool for the diversity sampling stage. Five augmented instances provide sufficient structural variation for representative sampling while keeping augmentation and LLM inference costs low.
To avoid repeated LLM calls, \tool maintains a semantic augmentation cache. Generated values are stored and reused whenever the corresponding semantic category appears in future log groups.

\begin{table*}[!t]
\centering
\caption{Comparison of log parsers on rare log groups.}
\resizebox{\linewidth}{!}{
\begin{tabular}{l|cccc|cccc|cccc|cccc|cccc|cccc}
\toprule
\multirow{3}{*}{\textbf{Dataset}}
& \multicolumn{4}{c|}{\textbf{Rule-based}}
 & \multicolumn{4}{c|}{\textbf{Tree-based}}
 & \multicolumn{16}{c}{\textbf{LLM-based (Llama-3-8B)} }\\
 \cmidrule{2-17}\cmidrule{17-21}\cmidrule{21-25}
    & \multicolumn{4}{c|}{\textbf{AEL}}
    & \multicolumn{4}{c|}{\textbf{Drain}}
    & \multicolumn{4}{c|}{\textbf{LILAC-8shots}}
    & \multicolumn{4}{c|}{\textbf{LibreLog}}
    & \multicolumn{4}{c|}{\textbf{EFParser}}
    & \multicolumn{4}{c}{\textbf{\tool}}\\
    
    & PA & GA & FTA & FGA
    & PA & GA & FTA & FGA
    & PA & GA & FTA & FGA
    & PA & GA & FTA & FGA
    & PA & GA & FTA & FGA
    & PA & GA & FTA & FGA \\
    \midrule
    
    Apache 
    & 0.00 & \textbf{1.00} & 0.00 & \textbf{1.00}
    & 0.00 & \textbf{1.00} & 0.00 & \textbf{1.00}
    & 0.50 & \textbf{1.00} & 0.50 & \textbf{1.00}
    & 0.50 & \textbf{1.00} & 0.50 & \textbf{1.00}
    & 0.50 & \textbf{1.00} & 0.50 & \textbf{1.00}
    & \textbf{1.00} & \textbf{1.00} & \textbf{1.00} & \textbf{1.00} \\
    
    BGL

     & 0.20 & 0.83 & 0.22 & 0.71
     & 0.13 & 0.96 & 0.19 & 0.94
     & 0.52 & 0.86 & 0.61 & 0.88
     & 0.41 & 0.96 & 0.40 & 0.94
     & 0.60 & \textbf{0.96} & 0.64 & \textbf{0.94} 
    & \textbf{0.70} & \textbf{0.96} & \textbf{0.65} & \textbf{0.94}\\
    
    Hadoop
    & 0.46 & 0.92 & 0.33 & 0.91
    & 0.29 & \textbf{1.00} & 0.19 & \textbf{1.00}
    & 0.43 & 0.88 & 0.46 & 0.84
    & 0.60 & \textbf{1.00} & 0.58 & \textbf{1.00}
    & 0.76 & 0.97 & 0.73 & 0.95
    & \textbf{0.82} & \textbf{1.00} & \textbf{0.83} & \textbf{1.00}\\
    
    HDFS
    & 0.80 & \textbf{1.00} & 0.75 & \textbf{1.00}
    & 0.00 & \textbf{1.00} & 0.00 & \textbf{1.00}
    & 0.00 & \textbf{1.00} & 0.00 & \textbf{1.00}
    & 0.80 & \textbf{1.00} & 0.75 & 0.00
    & \textbf{1.00} & \textbf{1.00} & \textbf{1.00} & \textbf{1.00}
    & \textbf{1.00} & \textbf{1.00} & \textbf{1.00} & \textbf{1.00} \\
    
    HealthApp
    & 0.48 & \textbf{1.00} & 0.25 & 0.82 
    & 0.48 & \textbf{1.00} & 0.90 & \textbf{1.00}
    & 0.91 & \textbf{1.00} & \textbf{0.92} & \textbf{1.00}
    & 0.88 & \textbf{1.00} & 0.75 & 1.00
    & 0.88 & \textbf{1.00} & 0.76 & \textbf{1.00}
    & \textbf{0.92} & \textbf{1.00} & 0.84 & \textbf{1.00}\\
    
    HPC
     & 0.03 & \textbf{1.00} & 0.06 & 0.89
     & 0.038 & \textbf{1.00} & 0.11 & \textbf{1.00}
     & 0.07 & 0.23 & 0.09 & 0.27
    & 0.65 & \textbf{1.00} & 0.57 & \textbf{1.00}
    & 0.80 & \textbf{1.00} & 0.78 & \textbf{1.00}
    & \textbf{0.96} & \textbf{1.00} & \textbf{0.92} & \textbf{1.00} \\
    
    Linux
     & 0.27 & 0.79 & 0.27 & 0.87
     & 0.29 & \textbf{0.90} & 0.28 & \textbf{0.95}
     & 0.08 & 0.10 & 0.16 & 0.24
    & 0.70 & \textbf{0.90} & 0.67 & 0.89
    & 0.64 & 0.85 & 0.67 & 0.92
    & \textbf{0.76} & 0.84 & \textbf{0.86} & 0.87 \\
    
    Mac
     & 0.16 & 0.92 & 0.16 & 0.90
     & 0.13 & 0.91 & 0.14 & 0.93
     & 0.29 & 0.85 & 0.44 & 0.93
     & \textbf{0.48} & 0.91 & 0.43 & 0.92
    & \textbf{0.48} & \textbf{0.96} & 0.48 & \textbf{0.96}
    & 0.46 & 0.93 & \textbf{0.45} & 0.92\\
    
    OpenSSH
    & 0.27 & \textbf{1.00} & 0.37 & \textbf{1.00}
    & 0.27 & \textbf{1.00} & 0.40 & \textbf{1.00}
    & 0.11 & 0.50 & 0.28 & 0.71
    & 0.44 & 0.61 & 0.51 & 0.80
    & 0.83 & \textbf{1.00} & 0.87 & \textbf{1.00}
    & \textbf{1.00} & \textbf{1.00} & \textbf{1.00} & \textbf{1.00}\\
    
    OpenStack
      & \textbf{0.66} & \textbf{1.00} & \textbf{0.66} & \textbf{1.00}
      & 0.33 & \textbf{1.00} & 0.33 & \textbf{1.00}
      & \textbf{0.66} & \textbf{1.00} & \textbf{0.66} & \textbf{1.00}
      & \textbf{0.66} & \textbf{1.00} & 0.66 &\textbf{1.00}
      & 0.33 & \textbf{1.00} & 0.33 & \textbf{1.00}
      & \textbf{0.66} & \textbf{1.00} & \textbf{0.66} & \textbf{1.00}\\
    
    Spark
    & 0.21 & \textbf{1.00} & 0.15 & 0.81
    & 0.17 & \textbf{1.00} & 0.17 & 1.00
    & 0.41 & 0.90 & 0.44 & 0.93
    & \textbf{0.46} & \textbf{1.00} & 0.45 & 0.30
    & \textbf{0.46} & \textbf{1.00} & \textbf{0.52} & \textbf{1.00}
    & 0.44 & \textbf{1.00}& \textbf{0.52} & \textbf{1.00}\\
    
    Thunderbird
     & 0.19 & \textbf{0.97} & 0.19 & 0.89
     & 0.17 & 0.95 & 0.18 & 0.94
     & 0.37 & 0.80 & 0.36 & 0.85
     & 0.53 & 0.86 & 0.48 & 0.90
     & 0.59 & \textbf{0.97} & 0.61 & \textbf{0.97}
    & \textbf{0.68} & 0.95 & \textbf{0.67} & 0.95\\
    
    Zookeeper
     & 0.44 & \textbf{1.00} & 0.50 & 0.80
     & 0.44 & \textbf{1.00} & 0.57 & \textbf{1.00}
     & 0.66 & \textbf{1.00} & 0.71 & \textbf{1.00}
     & 0.66 & \textbf{1.00} & 0.71 & \textbf{1.00}
     & 0.88 & \textbf{1.00} & 0.85 & \textbf{1.00}
     & \textbf{1.00} & \textbf{1.00} & \textbf{1.00} & \textbf{1.00}\\
    \midrule
    \textbf{Average}
     & 0.32 & 0.95 & 0.30 & 0.89
      & 0.21 & 0.98 & 0.27 & 0.98
     & 0.39 & 0.78 & 0.43 & 0.83
    & 0.60 & 0.93 & 0.57 & 0.83
     & 0.67 & \textbf{0.98} & 0.67 & \textbf{0.98}
    & \textbf{0.80} & \textbf{0.98} & \textbf{0.80} & \textbf{0.98}
    \\
\bottomrule
\addlinespace
\multicolumn{25}{l}{\textbf{Note:} The highest value in each metric is shown in bold. The Proxifier dataset is excluded because it does not contain any rare log groups.}

\end{tabular}
}

\label{tab:parser_comparison-rare}
\end{table*}

\subsection{Diversity Sampling}
The diversity sampling stage selects representative log instances from each log group before template inference. For rare log groups, the candidate pool consists of both the original and augmented log instances. For frequent log groups, the candidate pool contains only the original log messages, as sufficient structural variation is already available. 
Rather than providing all candidate logs to the LLM, we select five representative log instances that maximize structural diversity while reducing redundancy. This keeps the prompt concise while exposing the LLM to a wider range of variable representations.
To identify diverse examples, we compute the pairwise token-level Jaccard similarity~\cite{leskovec2020mining} between candidate logs and prefer logs with low overlap. We adopt token-level Jaccard similarity because prior work~\cite{LibreLog2024} found it to be an effective metric for selecting structurally diverse log messages.
Since the logs are already tokenized, Jaccard similarity naturally captures shared structural tokens without requiring additional model inference. 
We iteratively select the log with the lowest similarity to the previously selected examples until five representative instances have been chosen.

\subsection{Template Inference and Validation}

The representative log instances selected by the diversity sampling stage are then used for template inference. Our goal is to identify the static tokens shared across the selected logs while abstracting the dynamic values using wildcard placeholders. To improve the consistency of the generated templates, we also validate the inferred template against all log messages in the corresponding log group.


\noindent\underline{\textit{Template Inference.}}
We perform template inference by prompting the LLM with representative logs selected from the candidate pool. These logs share the same structure but contain different variable values, which allows the LLM to distinguish between static tokens and dynamic variables.
The prompt contains three parts. First, we provide a task instruction that asks the LLM to generate one template that matches all input logs. Second, we include an input-output example to standardize the expected response format. Third, we provide the selected representative logs and instruct the LLM to replace dynamic values with \texttt{<*>} while preserving static tokens unchanged.
By using representative logs from the augmented candidate pool, we expose the LLM to more variable values than the original rare group alone. This additional variation helps the LLM avoid treating rare variable values as static tokens, while the template-preserving augmentation ensures that the inferred template remains consistent with the original log structure.

\noindent\underline{\textit{Template Validation.}}
Although augmentation provides more structural evidence, the LLM may still generate an incorrect template. For example, it may preserve a dynamic value as a static token or replace a fixed token with a wildcard. To reduce such errors, we validate each inferred template before accepting it.
Specifically, we convert the template into a regular-expression pattern by replacing each \texttt{<*>}  placeholder with a wildcard-matching expression. We then check whether the template matches every original log instance in the corresponding group. If the template matches all logs, it is accepted and passed to the caching stage.
If validation fails, we select another representative subset from the candidate pool and query the LLM again. This allows us to infer the template from a different combination of examples. To keep the cost bounded, we limit this iterative validation process to three attempts. If no template passes validation after three attempts, we return the template from the final attempt.
This validation step provides a lightweight safeguard against occasional LLM errors. Rather than trusting the generated template directly, we check whether the template can explain the full log group.

\noindent\underline{\textit{Template Caching}}
Once a template passes validation, we store it in a template cache for future reuse. When a new log group is formed, we first check whether its template has already been inferred and stored in a template cache. If a matching template is found, we reuse the cached template directly without invoking the remaining stages. Otherwise, we determine whether the log group is rare. Rare log groups first go through template-preserving augmentation to enrich the available structural evidence, whereas frequent log groups proceed without augmentation. The resulting candidate logs are then passed to the diversity sampling stage, followed by template inference and validation. Once a template has been successfully validated, it is added to the cache for future reuse. This design avoids redundant LLM inference while preserving overall parsing performance.

\section{Study Results}
\subsection{RQ1: How effective is \tool on rare logs?}\label{sec:rare_log}

\phead{Motivation.}
Rare log groups remain challenging for existing parsers due to limited structural evidence. In this RQ, we evaluate whether \tool improves parsing performance on these groups.

\phead{Approach.}
We compare \tool against the baselines introduced in Section~\ref{sota_performance} on the rare log groups in Loghub-2.0 using the same experimental settings and evaluation metrics.

\phead{Results.}
Table~\ref{tab:parser_comparison-rare} reports the performance of all parsers. 

\noindent\textbf{\tool consistently balances accurate template inference and log grouping on rare log groups.} Compared with the strongest baseline, EFParser, \tool improves PA and FTA by 19\%, while maintaining similar GA and FGA.
Most baselines achieve relatively high GA but much lower PA. This indicates that they can group rare logs correctly but often fail to infer the corresponding templates. 
Unlike existing LLM-based parsers, which infer templates directly from log groups with limited structural evidence, \tool first enriches rare log groups through template-preserving augmentation. The additional structural evidence helps the LLM distinguish static tokens from dynamic variables and infer more accurate templates.
The improvements are also reflected in the template-level metrics. Unlike PA and GA, which evaluate individual log messages, FTA and FGA evaluate the correctness of the inferred templates themselves. This is particularly important for rare log groups, where incorrect template extraction may obscure abnormal or failure-related events, even if the associated log messages are grouped correctly. The strong template-level performance of \tool indicates that it not only groups rare logs accurately but also recovers the underlying template structure reliably.

To assess statistical significance, we conduct a Friedman test~\cite{Friedman1937}, followed by one-sided Wilcoxon signed-rank tests~\cite{Wilcoxon1945} comparing \tool against each baseline parser. The Friedman test indicates significant performance differences among the evaluated parsers across all metrics ($p < 0.01$). The pairwise Wilcoxon tests show that \tool significantly outperforms the other baselines on PA and FTA, with large effect sizes. 

\begin{tcolorbox}{\textbf{Answer to RQ1}: 
\tool effectively addresses data imbalance by significantly improving parsing performance on rare log groups. It achieves the highest PA and GA while maintaining state-of-the-art GA and FGA.}
\end{tcolorbox}

\subsection{RQ2: What is the impact of each component in \tool?}

\begin{table}[!t]
\centering
\caption{Ablation study of \tool components.}
\label{tab:seplog_ablation}
\resizebox{\columnwidth}{!}{
\begin{tabular}{lllll}
\toprule
\textbf{\tool} & \textbf{PA} & \textbf{GA} & \textbf{FTA} & \textbf{FGA} \\
\midrule
w/o log augmentation & 0.68 (15\%\drop{↓}) & 0.96 (1\%\drop{↓}) & 0.64 (20\%\drop{↓}) & 0.95 (1\%\drop{↓})\\
w/o semantic template & 0.60 (25\%\drop{↓})& 0.97 (-) & 0.55 (31\%\drop{↓}) & 0.96 (1\%\drop{↓}) \\
w/o template validation  & 0.59 (26\%\drop{↓}) & 0.96 (1\%\drop{↓}) & 0.57 (28\%\drop{↓}) & 0.96 (1\%\drop{↓}) \\
\bottomrule
\end{tabular}
}

\end{table} 
 
\phead{Motivation.}
\tool improves template inference through several complementary components. In this RQ, we perform an ablation study to quantify the contribution of each component.

\phead{Approach.}
We perform an ablation study by removing one component of \tool at a time. Specifically, we evaluate \tool without log augmentation, semantic template recognition, or template validation using PA, GA, FTA, and FGA.

\phead{Results.} 
Table~\ref{tab:seplog_ablation} summarizes the impact of each component.

\noindent\textbf{\tool's components primarily improve template recovery, with template validation and semantic template generation contributing the largest gains.}
Among the evaluated components, removing template validation causes the largest performance drop, reducing PA and FTA by 26\% and 28\%, respectively. Removing semantic template generation also significantly decreases performance (25\% in PA and 31\% in FTA), while removing log augmentation reduces PA and FTA by 15\% and 20\%, respectively. In contrast, GA changes only marginally across all settings, which suggests that the components primarily improve template recovery rather than log grouping. Overall, our findings show that log augmentation, semantic template generation, and template validation provide complementary benefits for template recovery under long-tailed log distributions.


\begin{tcolorbox}{\textbf{Answer to RQ2: }}{The ablation study confirms that log augmentation, semantic template generation, and template validation each make complementary contributions to \tool. }
\end{tcolorbox}
\subsection{RQ3: Does \tool generalize across different LLMs and LLM-based parsers?}

\phead{Motivation.}
Template-preserving augmentation (TPA) is designed as a modular framework that enriches rare log groups before template inference and can be integrated into existing LLM-based parsers without changing their core designs. Moreover, because \tool uses an LLM for both augmentation and template inference, we evaluate whether its effectiveness depends on a specific LLM backbone. Therefore, we investigate whether \tool generalizes across different LLMs and whether TPA consistently improves existing LLM-based parsers.

\phead{Approach.}
We conduct two experiments. First, we evaluate \tool with three different LLM backbones: Llama-3-8B, Mistral-7B, and Gemma-2-9B. We select these models because they are widely used, publicly available, and have comparable inference costs while differing in architecture and reasoning capability.
Second, we integrate TPA into EFParser, LibreLog, and LILAC-8shots while keeping their original parsing pipelines unchanged.
\begin{table}[!t]
\centering
\caption{Generalizability of \tool across LLMs and parsers.}
\label{tab:generalization}
\resizebox{\columnwidth}{!}{
\begin{tabular}{lllll}
\toprule
\textbf{Configuration} & \textbf{PA} & \textbf{GA} & \textbf{FTA} & \textbf{FGA} \\
\midrule
EFParser + TPA & 0.70 (4\%\gain{↑}) & 0.99 (2\%\gain{↑}) & 0.71 (5\%\gain{↑}) & 0.99 (1\%\gain{↑}) \\
LibreLog + TPA & 0.67 (14\%\gain{↑}) & 0.92 (1\%\drop{↓}) & 0.68 (16\%\gain{↑}) & 0.96 (17\%\gain{↑}) \\
LILAC-8shots + TPA & 0.55 (41\%\gain{↑}) & 0.93 (20\%\gain{↑}) & 0.55 (27\%\gain{↑}) & 0.94 (15\%\gain{↑}) \\
\midrule
TRAIL + Llama-3-8B & 0.80 & 0.97 & 0.80 & 0.97 \\
TRAIL + Mistral-7B & 0.70 (12\%\drop{↓}) & 0.92 (5\%\drop{↓}) & 0.77 (3\%\drop{↓}) & 0.93 (4\%\drop{↓}) \\
TRAIL + Gemma-2-9B & 0.80 (-) & 0.99 (2\%\gain{↑}) & 0.80 (-) & 0.98 (1\%\gain{↑}) \\
\bottomrule
\\
\multicolumn{5}{p{1.2\linewidth}}{\footnotesize
\textbf{Note:} For EFParser+TPA, LILAC-8shots+TPA, and LibreLog+TPA, percentages represent the relative performance change with respect to the corresponding original parser. For TRAIL+Mistral-7B, and TRAIL+Gemma-2-9B, percentages represent the relative performance change with respect to \tool using Llama-3-8B.
}
\end{tabular}
}
\end{table}

\phead{Results.}
Table~\ref{tab:generalization} summarizes the generalizability of \tool from two perspectives: different LLM backbones and existing LLM-based parsers. Overall, the proposed template-preserving augmentation generalizes well across both settings.

\noindent\textbf{Augmenting rare log groups with template-preserving examples consistently improves existing LLM-based log parsers.}
When integrated into EFParser, LibreLog, and LILAC-8shots without modifying their original designs, template-preserving augmentation improves PA by 4\%, 14\%, and 41\%, respectively. Similar improvements are observed for FTA and FGA. These results suggest that template-preserving augmentation improves existing LLM-based parsers by providing additional structural evidence, without changing their original designs.

\noindent\textbf{\tool maintains strong performance across different LLM backbones, indicating that its effectiveness is largely independent of the underlying model.} 
Replacing Llama-3-8B with Mistral-7B or Gemma-2-9B results in only modest performance differences. Gemma-2-9B performs slightly better than Llama-3-8B, whereas Mistral-7B shows a moderate decline across the evaluation metrics. Nevertheless, \tool maintains consistently high performance across all evaluated LLM backbones, suggesting that its effectiveness is largely independent of the underlying model.

\begin{tcolorbox}{\textbf{Answer to RQ3:}}
Template-preserving augmentation complements existing LLM-based parsers, while TRAIL maintains strong performance across different LLM backbones.
\end{tcolorbox}
\subsection{RQ4: How effective is \tool on the entire benchmark?}

\phead{Motivation.} 
Although \tool is designed to improve rare log parsing, deployments operate on complete datasets containing both frequent and rare log groups. In this RQ, we evaluate whether \tool maintains competitive performance on the complete datasets.

\phead{Approach.} 
We evaluate \tool on the complete Loghub-2.0 benchmark across all 14 datasets, which contains over 50 million log messages, using the same baseline parsers, experimental settings, and evaluation metrics described in Section 2.2. Unlike RQ1, this evaluation includes both frequent and rare log groups to assess whether TRAIL maintains competitive performance and robustness on complete datasets.

\begin{table}[!t]
\centering
\caption{Comparison of log parsers on the Loghub-2.0 benchmark.}
\label{tab:avg_results}
\resizebox{0.58\columnwidth}{!}{
\footnotesize
\begin{tabular}
{lrrrr}
\toprule
\textbf{Parser} & \textbf{PA} & \textbf{GA} & \textbf{FTA} & \textbf{FGA} \\
\midrule
AEL      & 0.39 & 0.80 & 0.29 & 0.57 \\
Drain    & 0.44 & 0.86 & 0.26 & 0.55 \\
LILAC-8shots & 0.69 & 0.85 & 0.54 & 0.77 \\
LibreLog & 0.79 & 0.85 & 0.64 & 0.81 \\
EFParser & 0.81 & 0.87 & 0.73 & \textbf{0.88} \\
\tool   & \textbf{0.86} & \textbf{0.90} & \textbf{0.74} & 0.85 \\
\bottomrule
\end{tabular}
}
\end{table}

\phead{Results.} 
Table~\ref{tab:avg_results} reports the average PA, GA, FTA, and FGA across the 14 datasets for each parser.

\noindent\textbf{Improving rare log parsing does not come at the expense of overall parser performance.} Despite being specifically designed to improve rare log parsing, \tool achieves the highest PA of 0.86, GA of 0.90, and FTA of 0.74 among all evaluated parsers. These results show that \tool improves the parsing of rare log groups without compromising performance on frequent ones. 
Although EFParser achieves a slightly higher FGA, \tool outperforms it on PA, GA and FTA. 
These results demonstrate that \tool is practical for real-world deployment, as its benefits extend beyond rare log groups to complete datasets containing both frequent and rare logs. 
To assess statistical significance, we conducted a Friedman test followed by one-sided Wilcoxon signed-rank tests. The Friedman test revealed significant differences among parsers for PA, FTA, and FGA ($p<0.05$), while the Wilcoxon tests showed that \tool significantly outperformed several baselines on PA and FTA and remained statistically comparable to EFParser.

\begin{figure}[!t]
    \centering
    \includegraphics[width=\linewidth, keepaspectratio]{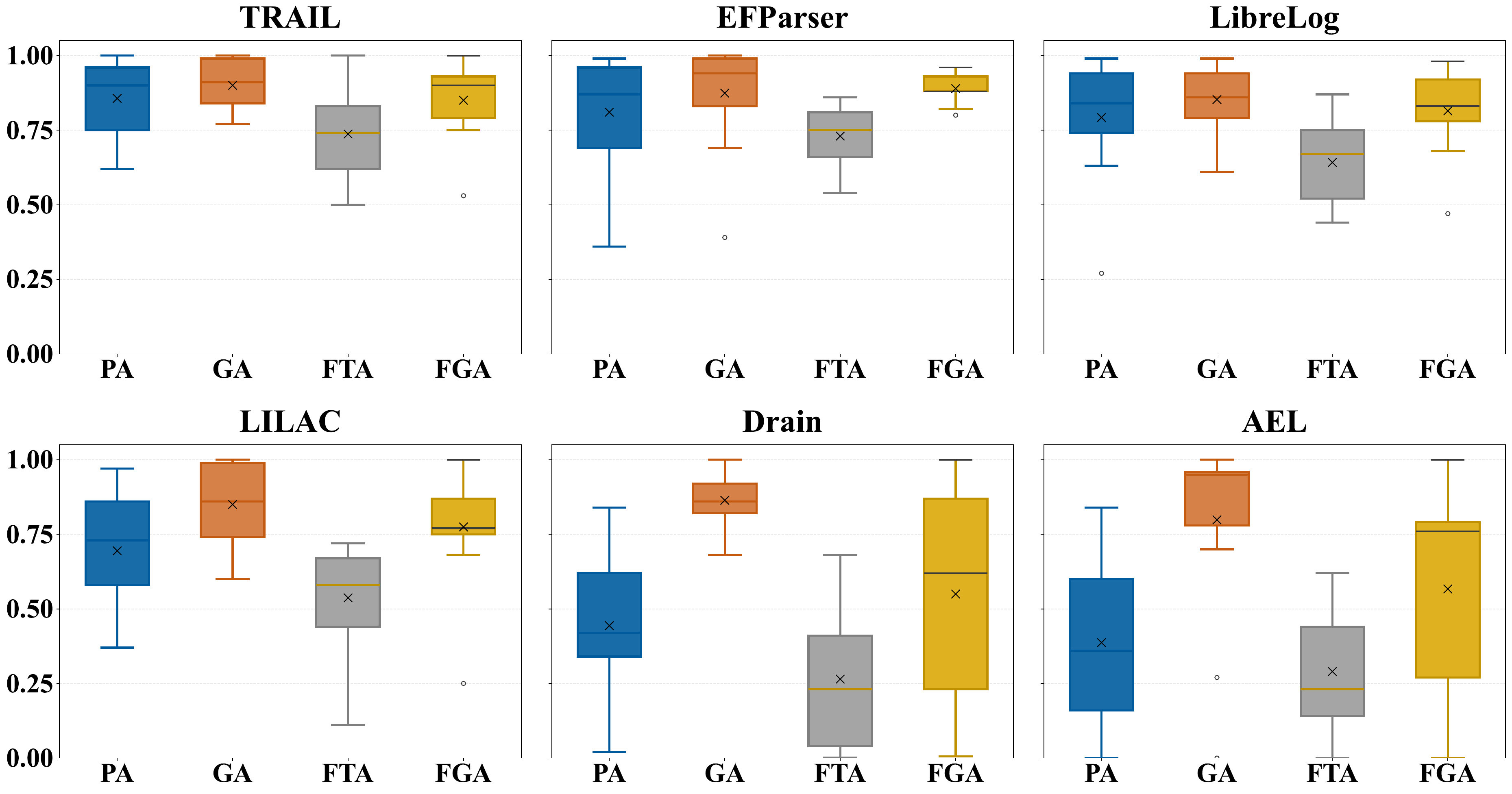}
    \caption{Robustness comparison between baselines and \tool on entire datasets.}
    \label{fig:boxplot}
\end{figure}

\noindent\textbf{\tool achieves robust performance across diverse logging systems.} Following prior studies~\cite{LILAC2024,zhu2019tools,cao2024robust,Uniparser2022,he2017towards}, we further analyze performance variance across datasets to assess parser robustness under diverse log distributions. 
As shown in Figure~\ref{fig:boxplot}, compared with the baseline parsers, \tool exhibits comparable or lower performance variation across the evaluation metrics.
For instance, its standard deviations are 0.13, 0.08, 0.13, and 0.12 for PA, GA, FTA, and FGA, respectively, compared with 0.19, 0.17, 0.10, and 0.05 for EFParser.
The lower variation in PA and GA indicates that \tool consistently maintains parsing and grouping performance across diverse datasets.
We attribute the overall robustness of \tool to the diversity sampling stage, which constructs a more representative candidate set for template inference.
By exposing the LLM to greater structural variation while avoiding redundant examples, \tool more consistently distinguishes static tokens from dynamic variables across datasets.
In contrast, the slightly higher variation in FTA and FGA suggests that recovering exact template boundaries remains inherently more sensitive to dataset-specific template complexity.

\begin{tcolorbox}{\textbf{Answer to RQ4}:}
\tool improves both overall effectiveness and robustness on the Loghub-2.0 benchmark. Despite targeting rare log parsing, it achieves the highest PA, GA, and FTA among all evaluated parsers.
\end{tcolorbox}

\subsection{RQ5: How efficient is \tool?}

\begin{table}[!t]
\centering
\caption{Average parsing time across Loghub-2.0 benchmark.}
\resizebox{.8\linewidth}{!}{
\begin{tabular}{lr}
\toprule
\textbf{Parser} & \textbf{Average Time (s)} \\
\midrule
AEL            & 4,891 \\
Drain          & 412  \\
LILAC-8shots   & 4,213 \\
LibreLog       & 3,086 \\
EFParser       & 1,503 \\
\midrule
\textbf{TRAIL} & 2,503 \\
 $\hookrightarrow$ Structural grouping          & 229 \\
 $\hookrightarrow$ Log augmentation             & 26 \\
 $\hookrightarrow$ Semantic template generation & 200 \\
 $\hookrightarrow$ Diversity sampling           & 2 \\
 $\hookrightarrow$ Template inference           & 1,373 \\
 $\hookrightarrow$ Template validation          & 673 \\
\bottomrule
\end{tabular}
}
\label{tab:time}
\end{table}










\phead{Motivation.} Efficiency is an important consideration for practical log parsing systems, particularly when processing large-scale production logs. In this RQ, we compare the efficiency of \tool with existing log parsers.

\phead{Approach.} We compare the average parsing time per dataset of \tool and the baseline parsers on the complete Loghub-2.0 benchmark. We also report the runtime breakdown of \tool's individual components. 
We conduct all experiments on identical hardware configurations, repeat each experiment three times, and report the average results to ensure reliability.

\phead{Results.} 
Table~\ref{tab:time} summarizes the runtime of all evaluated parsers on the complete Loghub-2.0 benchmark.

\noindent\textbf{\tool remains efficient in practice despite introducing additional processing stages.} Among the evaluated LLM-based parsers, EFParser achieves the shortest runtime at 1,503 seconds, followed by \tool at 2,503 seconds. In comparison, LibreLog and LILAC require 3,086 and 4,213 seconds, respectively. These results show that \tool maintains competitive efficiency while providing the additional processing needed to improve rare log parsing.

\noindent\textbf{The additional runtime of \tool is primarily spent on improving template inference rather than performing augmentation.} Template inference accounts for 1,373 seconds of the average runtime, while iterative validation requires 673 seconds. In contrast, structural grouping, semantic template generation, log augmentation, and diversity sampling together account for less than 20\% of the total execution time. These results indicate that the computational overhead of \tool is primarily associated with LLM inference, while the augmentation framework itself introduces only a modest cost.

EFParser is faster because it optimizes for a different design objective. EFParser emphasizes efficiency through cache-based optimization, whereas \tool performs additional template inference and validation to improve robustness on rare log groups. These additional stages increase runtime but enable more reliable template recovery under limited structural evidence. Future work may explore more efficient LLM-based template inference to further improve parsing efficiency.

\begin{tcolorbox}{\textbf{Answer to RQ5: }}{ Enriching structural evidence through augmentation is computationally inexpensive relative to LLM inference.}
\end{tcolorbox}

\section{Threat to validity}
\phead{External validity.}
Publicly available LLMs may have been pretrained on data that overlap with Loghub-2.0, which introduces the possibility of data leakage~\cite{LILAC2024, Ma2024LLMParser, LibreLog2024, wang2026small}. Although this threat cannot be completely eliminated, it applies equally to all evaluated LLM-based parsers, which use the same LLM backbones under identical experimental settings. 
It is also worth noting that the Loghub-2.0 evaluation benchmark with ground-truth templates was released in August 2023~\cite{jiang2024large}, whereas Llama-3-8B has a reported knowledge cutoff of March 2023~\cite{Llama3_2024}. Therefore, direct exposure of the benchmark during pretraining is unlikely.
Consequently, we expect the relative performance comparisons to remain largely unchanged even if some data leakage exists.

\phead{Internal validity.}
Since \tool relies on LLMs for template inference, nondeterministic outputs may influence the experimental results. To mitigate this threat, we set the temperature to 0 and repeat each experiment three times. We also evaluate multiple LLM backbones under identical settings.
\tool uses a lightweight set of regular-expression recognizers to identify common semantic categories during augmentation. The effectiveness of \tool may depend on how well these recognizers capture structured variables in the target logs. To mitigate this threat, the recognizers are used only to guide augmentation rather than to directly infer final templates, and unmatched tokens remain unchanged for template inference. 
In addition, our ablation study shows that removing semantic template generation reduces performance but does not make the technique inapplicable, suggesting that TRAIL is not solely dependent on the recognizer set. We also make the complete recognizer list publicly available to support reproduction and future extension.

\phead{Construct validity.}
We define rare log groups as those containing fewer than five instances. Although other thresholds are possible, we select five because it captures groups with insufficient structural evidence for reliable template inference while avoiding augmentation of groups that already provide sufficient evidence. We do not expect this design choice to substantially affect our overall conclusions, as \tool consistently improves parsing performance on the complete benchmark. Future work may investigate alternative threshold definitions and evaluate their impact on parser performance.

\section{Related work}

We categorize existing log parsers into two broad families: syntax-based and semantic-based methods.

\phead{Syntax-based parsers.} 
Traditional log parsers rely on heuristic rules and structural patterns to extract log templates~\cite{AEL2008,Drain2017,Logram2020,Spell2019}. For example, AEL~\cite{AEL2008} removes dynamic variables using heuristic rules, Drain~\cite{Drain2017} organizes logs using a fixed-depth parsing tree, Spell~\cite{Spell2019} applies longest common subsequence matching, and Logram~\cite{Logram2020} identifies frequent \textit{n}-gram patterns. These methods are efficient and effective for logs with recurring structural patterns. However, because they rely primarily on syntactic similarity, they often struggle to generalize to unseen log formats and complex log structures.

\phead{Semantic-based parsers.} Recent approaches leverage machine learning and language models to capture the semantic meaning of log messages~\cite{LILAC2024,Chen2024HighPrecision,Ma2024LLMParser,LibreLog2024,Xu2024DivLog,wang2026small,heng2025benchmarking,zhang2025logbase}. With the emergence of LLMs such as GPT~\cite{chatgpt2026} and Llama~\cite{Llama3_2024}, these methods generally achieve higher parsing accuracy than syntax-based approaches. 
For example, DivLog~\cite{Xu2024DivLog} and LLMParser~\cite{Ma2024LLMParser} leverage in-context learning to improve template inference. LILAC~\cite{LILAC2024} further improves efficiency through cache-based optimization, while LogPPT~\cite{Le2023PromptBased} employs a masked language model with few-shot learning to classify log tokens. LibreLog~\cite{LibreLog2024} advances this direction by combining LLMs with memory and iterative validation. More recently, EFParser~\cite{wang2026small} further enhances parsing through adaptive cache management and template validation.

Despite these advances, most semantic-based parsers are still evaluated primarily using overall parsing accuracy, which can mask failures on rare and underrepresented templates. Similar to many software engineering artifacts~\cite{song2018comprehensive,Zhou2023DevilInTheTails,han2025chase}, system logs exhibit a pronounced long-tail distribution, where a small number of frequent templates dominate while many templates occur only a handful of times. As a result, rare log groups contain too few examples for reliable template inference. To address this challenge, we propose \tool, a log parsing framework that enriches rare log groups with representative synthetic log instances before template inference. This additional evidence improves template recovery while preserving the original log semantics and structure.

\section{Conclusion}

In this paper, we introduced \tool, a log parsing framework designed to improve robustness under long-tailed log distributions. Through a large-scale empirical study on Loghub-2.0, we showed that rare log groups constitute a substantial portion of the template space despite representing only a tiny fraction of log messages. We further demonstrated that existing log parsers consistently struggle on these groups, highlighting the need for approaches that explicitly address limited structural evidence.
To address this challenge, \tool strengthens structural evidence for rare log groups through template-preserving augmentation before template inference. By combining semantic template generation, log augmentation, diversity sampling, and template validation, \tool enables more accurate identification of static tokens and dynamic variables. Our extensive evaluation shows that \tool consistently improves parsing accuracy on rare log groups while maintaining strong overall performance on complete datasets. 
We further demonstrate that the proposed augmentation strategy generalizes across different LLM backbones and can improve existing LLM-based parsers without modifying their core architectures.

\section{Data Availability}
We made our source code, prompts, datasets, and experimental results publicly available ~\cite{TRAIL_repo}.

\bibliographystyle{IEEEtran}
\bibliography{reference.bib}

\end{document}